\documentclass{sig-alternate}
\usepackage{amsmath}
\usepackage{times}
\usepackage{helvet}
\usepackage{courier}
\usepackage{graphicx}
\usepackage{color}
\usepackage{soul}
\usepackage{url}
\usepackage{xcolor}
\usepackage{stfloats}
\usepackage{caption}
\usepackage{authblk}
\title{Geo-Spotting: Mining Online Location-based Services for Optimal Retail Store Placement}

\newcommand{\affiliationfont}{\fontsize{8}{8}\selectfont}

\author[1]{Dmytro Karamshuk \thanks{This work was done while the author was based in the Computer Laboratory of the University of Cambridge.}}
\author[2]{Anastasios Noulas}
\author[2]{Salvatore Scellato}
\author[3]{Vincenzo Nicosia $^\ast$}
\author[2]{Cecilia Mascolo}
\affil[1]{\affiliationfont IMT Institute for Advanced Studies, Lucca, Italy}
\affil[2]{\affiliationfont Computer Laboratory, University of Cambridge, UK}
\affil[3]{\affiliationfont School of Mathematical Sciences, Queen Mary University of London, UK}

\begin{document}
\maketitle

\begin{abstract}
The problem of identifying the optimal location for a new retail store has been the focus of past research, 
especially in the field of land economy, due to its importance in the success of a business. Traditional approaches 
to the problem have factored in demographics, revenue and aggregated human flow statistics from nearby or remote areas. 
However, the acquisition of relevant data is usually expensive. With the growth of location-based social networks, 
fine grained data describing user mobility and popularity of places has recently become attainable.

In this paper we study the predictive power of various machine learning features on the popularity of retail 
stores in the city through the use of a dataset collected from Foursquare in New York. The features we mine are 
based on two general signals: geographic, where features are formulated according to the types and density of nearby 
places, and user mobility, which includes transitions between venues or the incoming flow of mobile users from distant areas. 
Our evaluation suggests that the best performing features are common across the three different commercial chains 
considered in the analysis, although variations may exist too, as explained by heterogeneities in the way retail 
facilities attract users. We also show that performance improves significantly when combining multiple 
features in supervised learning algorithms, suggesting that the retail success of a business may depend on multiple factors.

\end{abstract}




\section{Introduction}
\label{sec:intro}
The geographical placement of a retail store or a new business has been of prime importance
from the establishment of the first urban settlements in ancient times to today's modern trading and 
commercial ecosystems in cities. 
Open a new coffee shop in one street corner and it may thrive with hundreds of customers. 
Open it a few hundred meters down the road and it may close in a matter of months.

In this paper we take advantage of the new layers of information offered through check-in data in Foursquare
to frame the problem of optimal retail store placement in the context of location-based 
social networks. That is, given a set of candidate areas 
in the city to open a store, 
our aim is to identify the most promising ones in terms of their prospect to attract a large number of check-ins (i.e, become popular). 
We formulate this problem as a data mining task, where, by extracting a set of features, we seek to exploit them to assess
the retail quality of a geographic area.
In more detail, our contributions are:
\newline

\textbf{Spatial and mobility analysis of retail store popularity.} We conduct an analysis of the popularity of Foursquare venues: these follow a power-law distribution,  
which indicates the existence of very heterogeneous check-in patterns across places. Focusing on the venues of three retail stores chains we explore how their popularity
is shaped by spatial and human mobility factors. We find that $50$\% of user movements originate from nearby venues within $200$ to $300$ meters, and $90$\% of movements occur
within $1$km.
This suggests a strong \textit{local} bias in the attraction of customer crowds. Moreover, the geographic placement of retail stores with respect to different venue types is
non-random: there is, in general, a higher likelihood of observing a store near a transportation hub or a touristic spot (museum, hostel, etc.), compared to a randomly picked Foursquare venue type. 
We subsequently extend this analysis by considering the movements between retail chains and other places in proximity, discovering that co-location does not necessarily imply higher
probability of movement between two types of venues. This highlights how a deeper insight into human mobility patterns captured through location-based services in the city can improve the performance of static prediction frameworks on local business analytics.
\newline

\textbf{Mining features in location-based services for the retail assessment of a geographic area.}
Driven by the findings reported above we mine Foursquare user check-ins so as to capture a variety of signals that may be informative about the retail quality of a geographic area. The data mining features we choose belong 
to two well-defined classes: geographic and mobility. The \textit{geographic features} encode spatial information about the properties of
Foursquare venues in an area. This includes information about the existence of certain types of settlements (coffee shop, nightclub, etc.) in an area or its density.
Moreover, we factor the \textit{competitiveness} of an area by examining the influence of the presence of competitor business venues on the popularity of a retail facility. 
The set of \textit{mobility features} we devise involves the measurement of the popularity of an area in terms of number of check-ins: for instance, we model transition probabilities from nearby venues to the target store or the attraction of crowds from remote locations.

\textbf{Individual feature and supervised learning evaluation.} Finally, we assess the performance of the features mined in the previous step in terms of their ability 
to predict high rankings for the most popular geographic spots 
individually, and combined in a supervised learning framework.
Feature performance can vary across different retail chains as there are observed idiosyncrasies in the ways customer crowds are attracted at those. 
In principle, however, features accounting for the degree of competition in an area or those that model the spatial structure of an area, based on the existence of certain venue types, perform best in the ranking task. From the class of mobility features, the best predictors are those that rank areas based on their attraction of users from distant locations and those which consider the \textit{transition quality} of an area by means of venue types. 
When the features are combined using supervised learning models we observe a clear improvement in performance with the optimal retail spot being constantly ranked in the top-$5$\% of the prediction list in one over two cases. In addition, the supervised learning algorithms benefit marginally, yet steadily, when we employ mobility features in comparison to using those which encode solely the static geographic properties of an area. 

We envisage that similar approaches can influence research in  urban mining where a variety of applications may benefit, including the provision of better services for businesses and citizens or predictions of house price evolution, development indices of urban areas and location-based marketing. 
While Foursquare offers business accounts where shops can register themselves and obtain basic analytics such as the
times of the week their business is becoming more or less popular, or the demographics
of the users that check in to them~\cite{4sqMerchant}, data mining approaches that exploit the rich 
spatio-temporal datasets sourced from these services can considerably boost their business model by offering new commercial opportunities to their users
beyond their principal application scope so far that has been location-based activity recommendations~\cite{bao2012location}.

\section{Dataset Analysis}
\label{sec:analysis}
In this section we provide essential details about the Foursquare check-in dataset and subsequently 
we elaborate on the analysis of venue and retail store
popularity in the service. Particularly, we concentrate on the three most popular retail 
chains in the New York area and study their spatial and mobile interactions with 
other Foursquare venues and the mobility patterns of the users that visit them. 

\subsection*{Data Collection}
\label{subsec:collection}

Foursquare was launched in 2009 and it has quickly become the most popular
location-based service, with more than 35 million users as of January 2013~\cite{Foursquare2013}.
Per-user Foursquare check-in data is not directly accessible, however, users can
opt to share their check-ins publicly on Twitter.  We thus were able
to crawl for publicly-available check-ins via Twitter's streaming
API\footnote{\url{https://dev.twitter.com/docs/streaming-api}}. Note that we can
only access those check-ins that users explicitly chose to share on Twitter,
although users have the possibility to set this option as default. 
In the present work we use a dataset of check-ins and venue information in 
city of New York and the surrounding area. 
New York is the city where the service
was launched, and due to this fact the adoption of the service in the area is significantly larger than in any other
place in the world. Numerically speaking, we consider the square region of $10\times10$ km around the geographical center of 
New York (Manhattan Area, Coordinates: 40$^\circ$ 45'50''N 73$^\circ$ 58'48''W), featuring $37,442$ geo-tagged venues, $46,855$ users 
and $620,932$ check-ins collected in a period of 5 months (May 27th to November 2nd 2010). We note that according to our 
estimates this sample accounts for approximately $25\%$ of all check-ins collected by Foursquare in the aforementioned 
region and time frame. 

\begin{table}[!t]
\begin{tabular*}{\hsize}{@{\extracolsep{\fill}}rrrrr}
\hline
\multicolumn1c{Chain Name}&\multicolumn1c{Check-Ins}&
\multicolumn1c{Places}&
\multicolumn1c{$\langle c_p \rangle$}\cr
\hline
\multicolumn1c{Starbucks}&\multicolumn1c{$210,174$}&\multicolumn1c{$186$}&\multicolumn1c{$1129.97$}\cr
\multicolumn1c{Dunkin' Donuts}&\multicolumn1c{$26,955$}&\multicolumn1c{$104$}&\multicolumn1c{$259.18$}\cr
\multicolumn1c{McDonald's}&\multicolumn1c{$15,014$}&\multicolumn1c{$66$}&\multicolumn1c{$227.48$}\cr
\hline
\end{tabular*}
\caption{Summary of Chain Statistics: Total number of check-ins as observed through Foursquare's Venue API\protect\footnotemark \ at the time of data collection, number of places and average number of check-ins per place ($\langle c_p \rangle$)} 
\label{table:chainstats}
\end{table}

\footnotetext{\url{https://developer.foursquare.com/docs/venues/venues}}

\begin{figure}[ht]
\centering
\includegraphics[width=1.0\columnwidth]{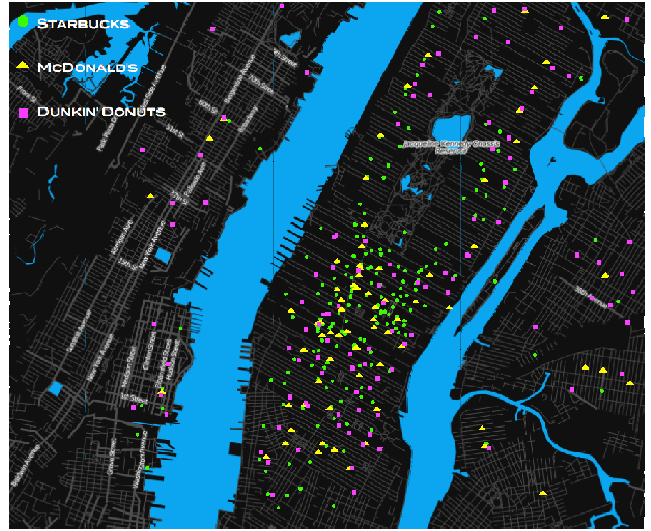}
\caption{Spatial distribution of the three store chains in the New York area.}
\label{ccdf_chains}
\end{figure}

\subsection*{Analysis of Retail Store Popularity}

\begin{figure}[ht]
\centering
\includegraphics[width=1.0\columnwidth]{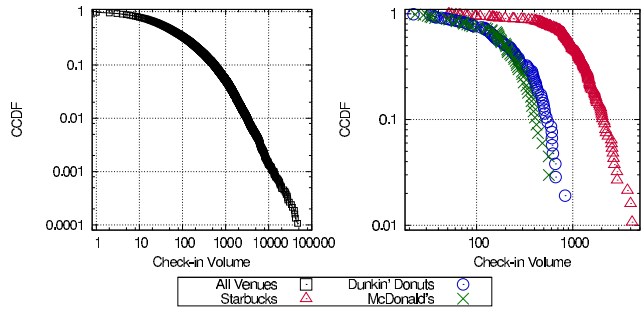}
\caption{Complementary Cumulative Distribution Function (CCDF) of check-ins per place for all venues in the dataset (left) and for the three considered store chains (right).}
\label{chainPop}
\end{figure}

Over the dataset just described we have analyzed the popularity of the different places present. 
By popularity of a place we mean the total number of check-ins we have observed in this venue. 
Figure~\ref{chainPop} illustrates the Complementary Cumulative Distribution
Function (CCDF) of check-ins per venue in the dataset, considering all places in New York.
The functional form of the distribution resembles a power-law 
and the check-in frequency spans a large number of orders of magnitude. 

From this point on we concentrate on the analysis of venues' popularity for 
individual store chains. We have observed considerable discrepancies between the check-in patterns across
different chains. In Table~\ref{table:chainstats} we show basic statistics for the three 
chain stores we have elected based on their number of venues in New York. 
Starbucks is the chain with the 
highest number of venues, $186$. Dunkin' Donuts and McDonald's follow with $104$ and $66$ stores, respectively. 
In Figure~\ref{ccdf_chains}, we present a visualization of the spatial distribution of the stores for the three retail chains we analyze.
The mean number of check-ins per place at Starbucks is equal to $1129.97$, almost five times larger
than the rest. Similarly, the CCDF of check-in volume for Starbucks restaurants, as shown in Figure~\ref{chainPop}, 
features a significantly longer tail than in the other two cases. Among other reasons, that could be attributed to the fact that visitors 
of a coffee shop stay there longer and thus are more likely to check-in. On the other hand, fast food chains may 
attract opportunistic visits just to pick up food. Addressing the problem thus for each chain separately can help us alleviate the biases
that may be introduced by check-in variations across place types in Foursquare.  

\subsection*{Spatial Interactions between Business Activities}
\label{sec:interactions}
Previous work aiming to assess the retail quality of places 
based on a dataset with similar spatial characteristics was 
proposed by Jensen in~\cite{jensen2006}. The author studied 
the spatial interactions between different place categories in the city of Lyon, France. 
The proposed metric computes the frequency of co-location 
for different dyads of place types, for instance, coffee shops close to railway stations, restaurants close to shopping centers,
and compares the results to the corresponding expectation when settlements are distributed uniformly at random in the city.
If the measured frequency is higher than the expected value, it is proposed that the categories 
tend to attract each other. If it is lower the two place types repel each other. 
The resulting inter-type coefficients are then exploited to assess the retail quality of an area
for a given place type.

Provided with the rich spectrum of place types (e.g., coffee shop, fast food restaurant, train station etc.)
that Foursquare venues can be semantically annotated with,
we compute the inter-category attraction coefficients for $248$ different venue categories (we discuss the details 
of the computation in Section~\ref{sec:jensen}).
The results of the Top 10 attractors for each chain with the corresponding 
inter-category coefficients are shown in Table~\ref{table:jensen}.
We observe that Starbucks and McDonald's restaurants are often placed near Train Stations  
with frequency $11.80$ (for Starbucks) and $3.08$ (for McDonald's) times higher than expected 
in the scenario when venues are distributed randomly in the city. This can be intuitively explained by the aspiration to benefit 
from the massive flows of people generated around transportation hubs. These crowds may seek food and refreshment opportunities
as they travel. Another common attractor for the two chains are Corporate Offices,  as these are good 
sources of regular clients. Similarly, the two coffee shop chains, Starbucks and Dunkin' Donuts, 
are often placed around Museums and Hostels are a source of tourist crowds. 
Less frequently, but, still considerably more often than expected at
random, Dunkin' Donuts can be found next to Gas Stations. 

\begin{table}\scriptsize
\centering
\scalebox{0.87}{
		\begin{tabular}{|ll|ll|ll|}
			\hline
			\textbf{Starbucks} 	&	& \textbf{Dunkin' Donuts} & & \textbf{McDonalds} & \\ \hline
			Train Station & \texttt{11.80}		 	&	Hostel 								&	\texttt{5.02} &		Flower Shop 		 & \texttt{5.87}			\\ 
			Light Rail    & \texttt{8.60} 				&	Gas Station 				 &		\texttt{3.05} &		Office Supplies & \texttt{3.16}			\\
			Stadium 			& \texttt{7.25}		 	  &	Automotive Shop 		&		\texttt{2.66} &		Train Station 	 &\texttt{3.08}			\\
			Airport 			& \texttt{6.24}		 	  & Flower Shop 				&		\texttt{2.36} &		Theater 				 &\texttt{2.84}			\\
			Museum 				& \texttt{5.10}		 	  & Post Office 				&		\texttt{2.19} &		Light Rail 			 &\texttt{2.32}			\\
			Convention Center  & \texttt{4.93} & Flea Market  			& \texttt{1.84} &		Gift Shop 			 &\texttt{2.26}			\\
			Hostel 						 & \texttt{4.82}	&	School 								&	\texttt{1.72} &		Subway Station 	 &\texttt{2.21}			\\
			Corporate Office & \texttt{4.57}	&	Drug Store 						 &	\texttt{1.70} &		Department Store &\texttt{2.17} 			\\
			Hotel 						& \texttt{4.13}	& Subway Station 				&	\texttt{1.67} &		Bank / Financial &\texttt{1.92}			\\
			Bank / Financial 	& \texttt{4.09}	& Bike shop 						 & \texttt{1.64}	&		Drug Store 			& \texttt{1.89}			\\ \hline
		\end{tabular} 
}		
	\caption{Top 10 most attractive categories for each chain as defined by Jensen's inter-category coefficients with the corresponding 
		values of the coefficients}
		\label{table:jensen}
\end{table}
	
\subsection*{Exploiting Mobility Data for Retail Analysis}
\begin{figure}[ht]
\centering
\includegraphics[width=0.75\columnwidth]{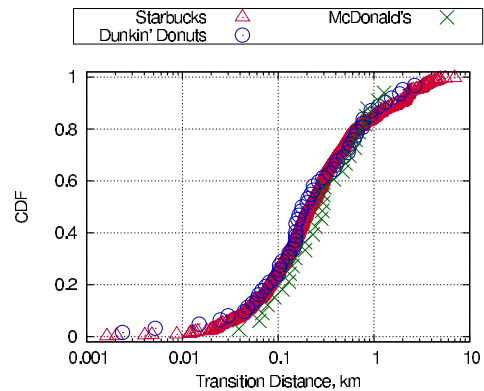}
\caption{The Cumulative Distribution Function (CDF) of the transition distances towards the shops of the three chains under analysis.}
\label{fig:transition_distance}
\end{figure}

Motivated by Jensen's approach~\cite{jensen2006} discussed above we extend the analysis of the retail quality 
of urban areas by also considering the fine grained information on human movements available in Foursquare. 
We will see how location-based social networks provide a unique opportunity to assess a geographic area 
not only by considering static spatial information, but also the dynamics of the movements of mobile users.  
In more detail, we study users' transitions between 
places inferred from their consecutive check-ins in different venues. As we show 
below, the aggregated transition data can be effectively exploited to analyze the flows of the users towards 
the place of interest and generally in the surrounding neighborhood. 
In Figure~\ref{fig:transition_distance} we plot the Cumulative Distribution Function (CDF) 
of incoming transition distances, i.e., distances that users travel from other venues in the dataset towards 
the three chains under analysis. 
As we observe in the plots, the vast majority of the incoming transitions, i.e., 
$80-90\%$, are done from within a radius of less than 1km and $50\%$ of transitions
from 200-300m. This result suggests that the customers of Starbucks coffee shops, Dunkin' Donuts 
and McDonald's restaurants usually come from local places. On the other hand the analysis of user transitions 
can also characterize interactions between Foursquare venues located at the different corners of the city:
this information is not captured by the place co-location approach discussed in the previous paragraph. 

We further measure the probability of transitions between different types of venues 
to determine the main donors of users towards the three chains under analysis (we discuss 
the details in Section~\ref{sec:transition}). The transition ratio $\rho$ is defined as the 
ratio between the transition probability of a given pair of venue categories over the 
random transition probability between any pair of categories.
The Top-10 place categories with the highest transition ratio towards the three chains under analysis
are presented in Table~\ref{table:transitions}. Numerous similarities are observed with respect to the 
inter-category attractiveness presented in Table~\ref{table:jensen}. The main 
attractors that we have identified by measuring the frequency of pairwise category co-location are often the main sources of customers. 
Thus, Hostels, Train Stations and Financial Centers, which had one of the highest attractiveness coefficients towards Starbucks restaurants, 
have also the highest transitional ratio towards them, i.e., 
$17.44$ (for Hostels), $4.79$ (for Train Stations) and $4.32$ (for Financial Centers) 
times higher than in a random scenario. 
The same observation holds for the attractiveness towards Schools and 
Subway Stations, in the case of Dunkin' Donuts. However, a deviation between the two measurements is also possible. 
The fact that a place type tends to be co-located with another may not necessarily mean that movement occurs across the two types of
places. 
Thus, considering information 
about user flows between venues can provide additional value 
to tackle the optimal retail store placement problem as will also become apparent during evaluation in Sections~\ref{sec:evaluation} and~\ref{sec:supervised}.

\begin{table}\scriptsize
\centering
\scalebox{0.87}{
		\begin{tabular}{|ll|ll|ll|}
			\hline
			\textbf{Starbucks} & & 				\textbf{Dunkin' Donuts} &	& \textbf{McDonalds} & \\ \hline
			Hostel & \texttt{17.44} 							  & Convention Center &  \texttt{8.18} 		& Parks \& Outdoor & \texttt{15.76} \\ 
			Flea Market / Fair &  \texttt{7.28}     & Laundry &  \texttt{5.51}							& Sculpture &  \texttt{6.10} \\ 
			Sculpture &  \texttt{6.80} 				      & Post Office &  \texttt{3.22}					& School &  \texttt{4.84} \\ 
			Drug Store &  \texttt{4.96} 						& Mall &  \texttt{2.20}									& Light Rail &  \texttt{3.94}  \\ 
			Train Station &  \texttt{4.79} 				  & Playground &  \texttt{1.49}						& Bus Station &  \texttt{3.40} \\ 
			Bank / Financial &  \texttt{4.32}	      & Drug Store &  \texttt{1.43}						& Post Office &  \texttt{2.70} \\ 
			Post Office &  \texttt{4.01}            &	Subway Station &  \texttt{1.30}				& Plaza / Square &  \texttt{2.20} \\ 
			Technopark &  \texttt{3.76}             & Bank / Financial &  \texttt{1.26}			& Bank / Financial &  \texttt{2.18} \\ 
			Admin. Building &  \texttt{3.64}				& School &  \texttt{1.18} 							& Airport &  \texttt{2.09} \\ 
			Convention Center &  \texttt{3.39}			&	Gas Station &  \texttt{1.09}  				& Theater &  \texttt{1.87} \\ \hline
		\end{tabular} 
}
	\caption{Categories with the highest transition ratio $\rho$ towards the three chains under analysis.}
	\label{table:transitions}
\end{table}

\section{Optimal Retail Store Placement in Location-based Services}
\label{sec:features}
In this section we formalize the problem of optimal retail store placement 
in the context of location-based social networks.
Our goal is to identify the best area amongst
a candidate set of potential areas for a new store to be opened. 
After formulating the problem, we define
and discuss the features we have mined from the Foursquare 
dataset to predict the best geographic spots by ranking geographic areas according to the predicted retail quality.

\subsection*{Problem Formulation}
\begin{figure}[ht]
\centering
\includegraphics[width=1.0\columnwidth]{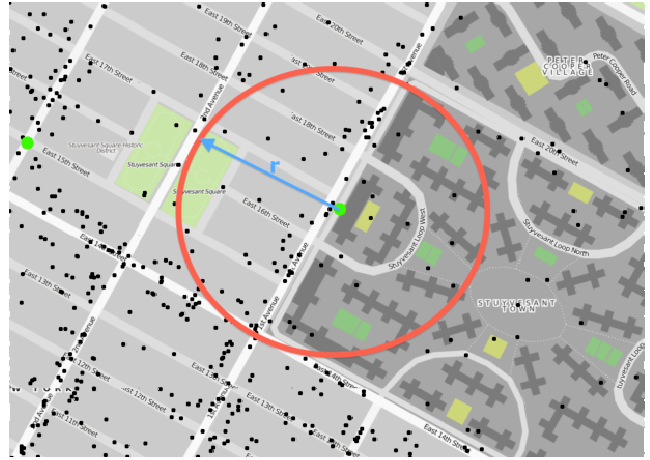}
\caption{Area of radius $r=200$ meters around a Starbucks coffee shop in New York. Black dots correspond to nearby Foursquare venues.}
\label{storesmap}
\end{figure}

Formally, by considering the existence of a candidate set of areas $L$ in which a commercial enterprise is interested in  
placing their business, we wish to identify the \textit{optimal} area $l \in L$, such that a newly open store in $l$ will potentially attract the largest number of visits. 
An area $l$ is encoded by its latitude and longitude coordinates 
and a radius $r$, as depicted in Figure~\ref{storesmap}. 
We have experimented with different values for the radius $r$ and have selected it to be equal 
to $200$ meters as this has yielded the best experimental results, i.e., the highest 
prediction performance across independent experiments with various values of $r$, 
and it is also in agreement with what the urban planning community considers 
as the optimal neighborhood size \cite{mehaffy2010}.
The ranking of places according to their popularity is then estimated using the features mined by incorporating the
characteristics of the area nearby (features are described next). For each feature we compute a score $\hat{\chi}_l$ for every candidate area $l$: 
the top ranked area in terms of that score will be the optimal area for the new store placement.
Our main assumption in the formulation of this task is that the number of empirically observed check-ins by Foursquare users can be used
as a proxy for the relative popularity of a place.

\subsection*{Prediction Features}

We now introduce the features we have mined from the Foursquare dataset in the city
of New York. Each feature returns a numeric score $\hat{\chi}_l$ that corresponds to a quality
assessment of the area for the optimal retail store placement problem. 
We have categorized the extracted features into two broad categories: 
\textit{place-geographic} features which integrate information about the types
and spatial interaction between the places, and \textit{user mobility} features 
which exploit knowledge about user movements and transitions between places. 

\subsubsection*{\textbf{Geographic Features}}
This class refers to features that describe the environment around the 
place under prediction as encoded through the spatial distribution
of Foursquare venues. More specifically, we measure the density, 
heterogeneity and competitiveness of the surrounding area by analyzing
the set $\{p \in P : dist(p, l) < r\}$ of places that lie in a disk of
radius $r$ around location $l$. The function $dist$ denotes the geographic
distance between two places and $P$ the set of venues in New York. 
\newline\newline
\textbf{Density:}
By measuring the number of neighbors around the place we assess
to what extent the popularity of a place depends on the density of other places 
in the same area. Formally:
\begin{equation*}\label{eq:neighbors}
	\hat{\chi}_l(r) = |\{p \in P : dist(p, l) < r\}|
\end{equation*}
We note that given that the radius $r$ is the same, the density 
of candidate geographic areas depends only on the number of places they include.
We denote the number of neighbors of a place within a radius $r$ with $N (l,r)$.
Intuitively, a denser area could imply higher likelihood for an opportunistic 
visit to a retail facility.
\newline\newline
\textbf{Neighbors Entropy:}
To assess the influence of the spatial heterogeneity of the area on the popularity of a place, 
we apply the entropy measure from information theory \cite{cover1991}
to the frequency of place types in the area. We denote the number of place neighbors 
of type $\gamma$ with $N_\gamma (l,r)$.
The entropy defines how many bits are required 
to encode the corresponding vector of type counters $\{N_\gamma (l,r) : \gamma \in \Gamma\}$, where $\Gamma$ is a set of all types,
and is higher the more heterogeneous the area is: 

\begin{equation*}\label{eq:entropy}
	\hat{\chi}_l(r) = -\sum_{\gamma \in \Gamma}{\frac{N_\gamma(l,r)}{N(l,r)} \times \log{\frac{N_\gamma(l,r)}{N(l,r)}}}
\end{equation*}
In general, an area with high entropy values is expected to be diverse in terms
of types of places, whereas the least entropic areas would imply that
the number of check-ins is biased towards a specific category, for instance \textit{Home} if the 
area is residential.
\newline\newline
\textbf{Competitiveness:}
We devise a feature to factor in the competitiveness of the surrounding area. 
Given the type of the place under prediction $\gamma_l$ (for example Coffee Shop for Starbucks), 
we measure the proportion of neighboring places of the same 
type $\gamma_l$ with respect to the total number of nearby places. 
We then rank areas in reverse order, assuming that 
the least competitive area is the most promising one:
\begin{equation*}\label{eq:competitors}
	\hat{\chi}_l(r) = - \frac{N_{\gamma_l} (l,r)}{N (l,r)}
\end{equation*}	
It is worth noting however that competition in the context of retail stores and marketing can 
have either a positive or a negative effect. One would expect that, for instance, 
placing a bar in an area populated by nightlife spots would be rewarded 
as there is already an ecosystem of related services and a crowd of people 
being attracted to that area. However, being surrounded by competitors may also mean 
that existing customers will be shared. 
\newline\newline
\textbf{Quality by Jensen:}
\label{sec:jensen}
To consider spatial interactions between different place categories, as we 
anticipated in Section~\ref{sec:interactions},
we exploit the metrics defined by Jensen et al. in \cite{jensen2006}. 
To this end, we use the inter-category coefficients described in the 
previous section to weight the desirability of the places observed 
in the area around the object, i.e., the more the places that attract the object
exist in the area, the better the quality of the location.
More formally, we define the quality of location for venue of type $\gamma_l$ as: 
\begin{equation*}\label{eq:jensen}
\hat{\chi}_l(r) = \sum_{\gamma_p \in \Gamma}{\log(\kappa_{\gamma_p \rightarrow \gamma_l}) \times (N_{\gamma_p} (l,r) - \overline{N_{\gamma_p}(l,r)})}
\end{equation*}
where $\overline{N_\gamma(l,r)}$ denotes how many venues of type $\gamma_p$ are observed on average around the places 
of type $\gamma_l$, $\Gamma$ is the set of place types, 
and $\kappa_{\gamma_l \rightarrow \gamma_p}$ are the inter-type attractiveness coefficients. To compute 
the latter, we analyze how frequently places of type $\gamma_l$ are observed around $\gamma_p$ on average, and normalize that 
value with the expectation for a random scenario. Formally we get: 
\begin{equation*}\label{eq:jensencoefficients}
	\kappa_{\gamma_p \rightarrow \gamma_l} = \frac{N - N_{\gamma_p}}{N_{\gamma_p} \times N_{\gamma_l}} \sum_p{\frac{N_{\gamma_l} (p,r)}{N (p,r) - N_{\gamma_p} (p,r)}}
\end{equation*}
where $N$, $N_{\gamma_l}$ and $N_{\gamma_p}$ denote the total number of places considered in the analysis and the number of 
places for types $\gamma_l$ and $\gamma_p$ respectively. Similarly, the intra-categories coefficient $\kappa_{\gamma_l \rightarrow \gamma_l}$ 
are computed between the places of the same type, thus, assessing to which extent the places tend to group into spatial clusters, e.g., 
financial centres at Wall Street.

\subsubsection*{\textbf{Mobility Features}}
In this section we show how information about the check-in patterns of Foursquare users
can be exploited to assess the retail quality of an area. Our goal is to identify to what
extent information crowdsourced from mobile users can improve geographic business analytics
and what are the benefits with respect to information exploiting only static spatial information 
about venues such as the features presented above.
We shall consider 
characteristics that measure the general popularity of the area and features that 
exploit transitions amongst venues. 
\newline\newline
\textbf{Area Popularity:}
To assess the influence of the overall popularity of the area on the popularity of 
individual places we measure the total number of check-ins empirically observed 
among the neighboring places in the area: 	
\begin{equation*}\label{eq:total}
\hat{\chi}_l(r) = |\{(m, t) \in C: dist(m,l) < r\}|
\end{equation*}
where tuple $(m, t)$ denotes a check-in recorded in place $m \in P$ at time $t$,
and $C$ is a set of all check-ins in the dataset.
\newline\newline
\textbf{Transition Density:}
Assuming that increased mobility between places in the area can increase 
the number of random visitors towards the target place, we measure the density 
of transitions between the venues inside the area. Formally, 
by denoting as $T$ the total set of consecutive check-in transitions between
places and  as a tuple, $(m,n) \in T$, the places $m \in P$ and $n \in P$
involved in two consecutive check-ins, we have: 
\begin{equation*}
\hat{\chi}_l(r) = |\{(m,n) \in T : dist(m,l) < r \wedge dist(n,l) < r\}|
\end{equation*}
\newline
\textbf{Incoming Flow:}
We also define a feature to account for the incoming flow 
of external user traffic towards the area of the place in question. 
We consider transitions between places denoted by a tuple, $(m,n) \in T$,
such that first place $m$ is located outside and second place $n$ inside the area under prediction. 
Formally: 
\begin{equation*}
\hat{\chi}_l(r) = |\{(m,n) \in T : dist(m,l) > r \wedge dist(n,l) < r\}|
\end{equation*}
One would expect that an area that is a good attractor of remote users would be a promising one for
establishing a new retail facility.
\newline\newline
\textbf{Transition Quality:}
\label{sec:transition}
Another aspect of location attractiveness comes from the potential number 
of local customers that the place might attract from the area. We measure 
the probability of transitions between all other types of venues and venues of the same 
type as the venue itself. The resulting probabilities allow us to weight the nearby places as 
potential sources of customers to the place under prediction. More formally:
\begin{equation*}
	\hat{\chi}_l(r) = \sum_{\{p \in P: dist(p, l) < r\}}{\sigma_{\gamma_p \rightarrow \gamma_l} \times C_p}
\end{equation*}
where $C_p$ is the number of check-ins at a place $p$ 
and $\sigma_{\gamma_p \rightarrow \gamma_l}$ is a probability of transitions between 
two categories $\gamma_p$ and $\gamma_l$, defined as the average percentage of 
all check-ins to place $p$ of category $\gamma_p$ 
that are followed by transitions to places of category $\gamma_l$: 
\begin{equation*}
	\sigma_{\gamma_p \rightarrow \gamma_l} = \mathrm{E}[\frac{|\{(m,n) \in T : m = p \wedge \gamma_n = \gamma_l\}|}{C_p}]
\end{equation*}

\begin{equation*}
	\rho_{\gamma_p \rightarrow \gamma_l} = \sigma_{\gamma_p \rightarrow \gamma_l} \frac{N - N_{\gamma_p}}{N_{\gamma_p} \times N_{\gamma_l}}
\end{equation*}

Having defined the two classes of geographic and mobility features, covering a diverse set of signals that can be exploited 
in location-based services, we evaluate their performance in the next sections aiming to understand what sources 
of information form the best predictors of the popularity of a retail chain unit.

\section{Feature Evaluation}
\label{sec:evaluation}
As discussed, a primary aim of this paper is to identify the 
most important information signals that can be mined from location-based services,
in order to forecast the area where a retail store may attract the maximum number 
of check-ins. To this end, we evaluate the performance of the
individual prediction features introduced in the previous section. 
We first describe the methodology and the metrics 
we employ; then we compare the performance of each individual feature
across these metrics. 

	\begin{figure}[htbp]
			\centering
			\includegraphics[width=1.0\columnwidth]{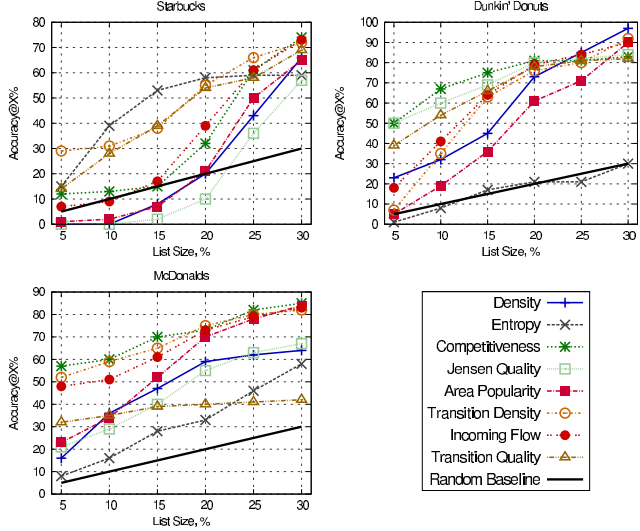}
			\caption{The Accuracy@X\% scores of the individual feature predictions for Starbucks, Dunkin' Donuts and McDonalds.}
			\label{fig:accuracy}
		\end{figure}

\subsection*{Methodology and Metrics}
Given a specific retailer brand like the three we are considering in this paper,
we analyze a set of prediction areas $L$ for a new store. 
For each feature and area $l \in L$ we compute $\chi_l$ as defined in the previous section and then rank the 
locations based on this.
As a result we obtain a ranked list of locations $R = (l_1, l_2, ..., l_{|L|})$
and denote with $rank(l_i)$ the position of location $l_i$ in $R$. 
We also compute the ranked list $\overline{R}$ of locations based on the actual popularity (number of check-ins)
of the stores in those locations and denote with 
$\overline{rank}(l)$ the position of location $l$ in $\overline{R}$. 
Given the two lists ranking locations in terms of predicted and ground truth values, 
we  formally define the metrics we use to assess
the quality of predictions achieved by the different features.
\newline

\textbf{NDCG for top-K location ranking:}
Firstly, we aim to measure the extent to which the top-k locations in the list of actual popularity 
$\overline{R}$ are highly ranked in the predicted list $R$. To this end, we adopt 
the NDCG@$k$ (Normalized Discounted Cumulative Gain) metric frequently used in the 
performance evaluation of information retrieval systems \cite{jarvelin2002cumulated}. The metric assesses
the cumulative gain achieved by placing the most relevant instances in the top-k 
of the prediction list as formally defined by the Discounted Cumulative Gain measure:
\begin{equation}
DCG@k = \sum_{i = 1}^{k}{\frac{2^{rel(l_i)}-1}{\log_2{(i+1)}}}
\end{equation}
where $rel(l_i)$ is the score relevance of an instance at position $i$ in the predicted ranking $R$.  
The result is then normalized by the DCG of the ideal prediction ($i$DCG),   
when the instances are sorted by the relevance with the most relevant in the first position. 
The resulting scores, thus, lie in the range from $0 < $ NDCG@$k \leq 1$. As a relevance 
score for an instance $l_i$ we use its relative position in the 
actual ranking $\overline{R}$, i.e., $rel(l_i) = \frac{|L| - \overline{rank}(l_i) + 1}{|L|}$.
The $rel(l_i)$ score is equal to $1$ when the area is ranked first in terms of check-ins 
and it linearly decreases to $0$ as the rank goes down the list. As a baseline for comparison, 
we use the expected value of NDCG@$k$ for a random ranker which is achieved by randomly 
permuting the instances in the testing set. 
\newline

\textbf{Assessing the best prediction:}
Considering now the application scenario where the best geographic area for a new business has to be discovered, for instance 
by a geo-analytics team, we would like to compare the different ranking strategies in terms of their ability to yield high quality 
top locations. To this end, we measure the fraction of times that the optimal location in the predicted list $R$ is at the top-X\% 
of the the actual popularity list $\overline{R}$ which represents our ground truth. We refer to this metric as Accuracy@X\% and we note
that we have used the \% instead of absolute list size values (i.e., top-K) to allow for comparison across different chains.   
\newline

\textbf{Geographic Cross Validation:}
We use a random sub-sampling method~\cite{maimon2010} to 
select subsets of geographic areas for validation.
In each experiment (repeated here $1000$ times) we randomly sample $33$\% of the total areas 
associated with the stores of the brand we analyze, obtaining $L$ candidate areas for prediction.
The rest of the areas are subsequently used to form our training set on which features (or supervised learning
algorithms as we see in Section~\ref{sec:supervised}) will be trained. We assume that the stores (and the corresponding 
check-ins) in the test set areas $L$ do not yet exist and our goal is to forecast the popularity ranking of the stores, 
if they were to be opened there. We note that the same assumption is used to define the features in the training and 
test sets. The mean NDCG@$k$ scores are measured by averaging across all testing sets. 

\begin{table}[htbp]
\centering
	\begin{tabular}{|c|c|c|c|}
		\hline
		\textbf{Feature} & \textbf{Starbucks} & \textbf{Dun.Don.} & \textbf{McDon.} \\ \hline
		\multicolumn{4}{|c|}{\textbf{Geographic Features}} \\ \hline
		Density 			 			& 0.60 & 0.79 & 0.73 \\ \hline
		Entropy 			 			& 0.65 & 0.60 & 0.69 \\ \hline
		Competitiveness 		& \textbf{0.70} & 0.68 & \textbf{0.78} \\ \hline
		Jensen Quality 			& 0.54 & \textbf{0.81} & 0.72 \\ \hline
		\multicolumn{4}{|c|}{\textbf{Mobility Features}} \\ \hline
		Area Popularity 		& 0.54 & 0.77 & 0.78 \\ \hline
		Transition Density 	& 0.62 & 0.79 & 0.78 \\ \hline
		Incoming Flow   		& 0.60 & 0.75 & \textbf{0.79} \\ \hline
		Transition Quality 	& \textbf{0.66} & \textbf{0.81} & 0.73 \\ \hline
		\multicolumn{4}{|c|}{\textbf{Random Baseline}} \\ \hline
		                    & 0.48 & 0.51 & 0.53 \\ \hline
	\end{tabular} 
	\caption{The Average NDCG@10 results of the individual feature predictions 
	for each of the three chains.}
	\label{table:individual_ndcg}
\end{table}

\subsection*{Individual Feature Performance}
\textbf{NDCG Top-K location ranking:}
In Table~\ref{table:individual_ndcg}, we present the results obtained for the NDCG@$10$ metric for all features across the three chains.  
In all cases we observe a significant improvement with respect to the random baseline, yet there are features which perform considerably better than others. 
In particular, with regards to the geographic class of features, the Jensen Quality does best with NDCG@$10$ = $0.81$ for Dunkin' Donuts,
whereas Competitiveness is the top feature for Starbucks and McDonald's scoring $0.70$ and $0.78$ respectively. 
Interestingly, as suggested by the Competitiveness feature, the lack of competitor venues in an area can have a positive influence in attracting customers. 
A potential interpretation of this observation is that an urban area, for instance a neighborhood, is expected to provide a number 
of services to its local residents. A retail facility can benefit by being the main provider at an area whereas, in contrast, its customer
share is expected to drop as more and more competitors are opening nearby.

When considering the class of mobility features, Transition Quality does best in the cases of Starbucks and Dunkin' Donuts with an NDCG@$10$ score of $0.66$ and $0.81$.
This signifies that knowledge about the types of places that can initiate large customer flows to a target business can be useful to the placement of a new shop.
In the case of McDonald's, however, the most effective feature is Incoming Flow which has achieved an NDCG@$10$ score of $0.81$. A careful inspection of Figure~\ref{fig:transition_distance} suggests
that this may be due to the fact that McDonald's tend to attract customers from remote locations with higher probability than the other chains. In the case of McDonald's the probability of attracting 
a Foursquare user from a distance beyond $200$ meters is almost $0.65$, when the value drops to $0.50$ and $0.45$ for Starbucks and Dunkin' Donuts respectively. 

Another observation which stands out is that in the case of Dunkin' Donuts, Jensen Quality and Transition Quality achieve the same performance (NDCG=$0.81$), better than the rest of the features. 
This is a case where geographic and mobility features agree in performance revealing that spatial structure alone is sufficiently correlated with the mobility of users in the area. A comparison 
between Tables~\ref{table:jensen} and~\ref{table:transitions} shows that the rankings of the different types of places across the three chains agree more for Dunkin' Donuts compared to Starbucks
and McDonald's, where larger deviations may signify a large discrepancy between the geographic and mobility properties of an area. 
\newline

\textbf{Assessing the best prediction:}
We now evaluate the individual feature predictors in terms of the Accuracy@$X$\% metric. As defined above, this metric considers the top predicted location $l \in L$ and measures how high it is ranked in the list of actual rankings $\overline{R}$. For small list fractions $5$\% the relative performance of the various features is qualitatively similar to the NDCG metric. While the ranking in the performance 
of different features persists across different list $X$\% values, it is worth mentioning that Competitiveness tends to rise faster than other features reaching a score of Accuracy@$15$\% higher than $70$\%, outperforming
the winning (in terms of NDCG) Incoming Flow for McDonald's and Jensen Quality for Dunkin' Donuts. The results in terms of Accuracy@$X$\% are much lower in the case of Starbucks. An explanation of the drop in performance (also observed for NDCG) for Starbucks may relate to their high density in the city: as shown in Table 1 (and hinted by observing Figure~\ref{ccdf_chains}), the number of Starbucks stores is almost two times higher than McDonald's and Dunkin' Donuts in the same $100$km$^{2}$ area around Manhattan.
Thus, the potential geographic overlap
of the areas covered by two Starbucks picked randomly is higher and, therefore, while the underlying features may be very similar their corresponding popularity may differ. To provide a numerical indication of this issue we note that the probability of encountering a store of the same chain within $100$ meters was $35$\% for Starbucks, whereas only $11$\% and $7$\% for McDonald's and Dunkin' Donuts, respectively.

\section{Supervised Learning Approach}
\label{sec:supervised}
	In this section we combine features in a supervised 
	learning framework. Our aim is to exploit the union of individual 
	features in order to improve predictions, testing if the popularity 
	of places in Location-based Social Networks can be better predicted by considering 
	a composition of signals. 
	We use different supervised models to learn how feature
	vectors $\mathbf{x}$ can be associated with the check-in scores $y$ of the areas under prediction. 
	Two different ranking methodologies are employed which we explain in the following paragraphs.
	
	\textbf{Supervised Regression for Ranking:}
	The three algorithms we exploit here are Support Vector Regression~\cite{hearst1998}, M5 decision trees~\cite{quinlan1992}
	and Linear Regression with regularization. The latter case assumes that 
	the output score $y$ is a linear combination of the input features $\mathbf{x}$ with the weights $\mathbf{w}$
	being calculated from the training data. The goal of the prediction 
	algorithm is to minimize the error between actual and predicted outputs:
	\begin{align}
		\underset{\textbf{w}} \min \Vert\mathbf{x}^{\mathsf{T}} \mathbf{w} - y \Vert^{2} + \gamma \Vert\mathbf{w}\Vert^{2}
	\end{align}
	where $\gamma$ is the regularization parameter set here equal to $10^{-8}$.
	We have used the corresponding implementations that are publicly available through the 
	WEKA machine learning framework \cite{witten2011}. By training the supervised learning algorithms to 
	obtain regression scores and, subsequently, rank the candidate geographic areas, we are essentially reducing 
	the regression problem to a ranking one. 
	\newline
	\begin{table}
	  \centering
		\begin{tabular}{|c|c|c|c|}
			\hline
			\textbf{Algorithm} & \textbf{Starbucks} & \textbf{Dun.Don.} & \textbf{McDon.}\\ \hline
			\multicolumn{4}{|c|}{\textbf{Geographic Features}} \\ \hline
			Lin. Regression 		& 0.73 & 0.80 & 0.78 \\ \hline
			M5 Dec. Trees 			& 0.72 & 0.80 & 0.78 \\ \hline
			SVR                 & 0.73 & 0.81 & 0.81 \\ \hline
			RankNet             & 0.72 & 0.81 & 0.79 \\ \hline
			\multicolumn{4}{|c|}{\textbf{All Features}} \\ \hline
			Lin. Regression 		& 0.76 & 0.82 & 0.82 \\ \hline
			M5 Dec. Trees 			& 0.77 & 0.82 & 0.82 \\ \hline
			SVR                 & 0.77 & 0.83 & 0.84 \\ \hline
			RankNet             & 0.77 & 0.81 & 0.83 \\ \hline
			\multicolumn{4}{|c|}{\textbf{Random Baseline}} \\ \hline
		                      & 0.48 & 0.51 & 0.53 \\ \hline
		\end{tabular} 
		\caption{The best Average NDCG@10 results of the supervised learning algorithms applied to combinations of features 
		grouped into two classes: combinations of geographic features only and combinations of both mobility and 
		geographic features.}
		\label{table:combination}
	\end{table}

\begin{figure}[ht]
	\centering
	\includegraphics[width=0.75\columnwidth]{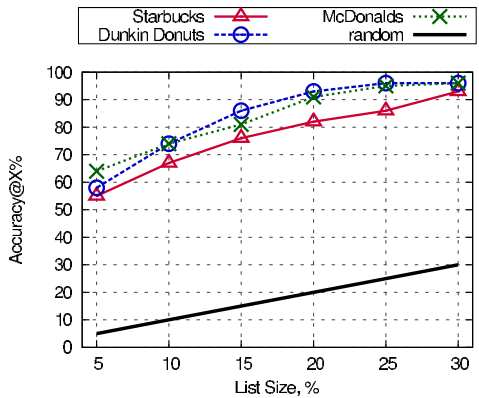}
	\caption{The Accuracy@X\% results of the best performing combination of features for the three store chains.}
	\label{fig:combination}
	\end{figure}

	\textbf{Supervised Learning to Rank:} Additionally, we consider a pair-wise learning-to-rank approach, RankNet \cite{burges2005learning},
	that learns the ordering relation between a pair of venues based on their features. Given a pair of
	venues $A$ and $B$, characterized by features $\mathbf{x^A}$ 
	and $\mathbf{x^B}$, RankNet identifies if venue $A$ has to be ranked 
	higher than $B$. The models assume a ranking 
	function $H(\mathbf{x}) : \Re^{|\mathbf{x}|} \rightarrow \Re$ such 
	that the rank order of instances is specified 
	by the real values of $H$. Specifically, 
	$H(\mathbf{x^A}) > H(\mathbf{x^B})$ assumes that 
	venue $A$ has to be ranked  higher than venue $B$. 
	As the ground truth in this model we employ a linear 
	ranking $\overline{R}$ of test set venues 
	according to the number of their check-ins. 
	We finally note that in the RankNet algorithm the ranking function $H(x)$ is 
	modeled as a neural network and we use a publicly available  
	implementation of the algorithm from the RankLib library\footnote{\url{http://people.cs.umass.edu/~vdang/ranklib.html}}.  
Overall, the testing and evaluation of the supervised 
algorithms is conducted using the exact methodology
we have used for the evaluation of individual features, and all features have been normalized 
before training.
	\newline
	\textbf{Results}
			We now present the results of supervised learning using the algorithms described above. 
			We measure the performance of the algorithms trained separately on different combinations of 
			geographic features and then combining them with mobility features. We aim to assess the
			extent to which the fusion of the features yields better results 
			than the prediction based on individual features and secondly, to what extent adding mobility 
			features improves predictability based only on geographical information. 
			The supervised models based on the grouping of geographic features offer a slight
			increase ($0.02$-$0.03$) in NDCG@$10$ performance in comparison to individual feature prediction 
			in two out of three cases: McDonald's and Starbucks. However, adding mobility features to the supervised models 
			considerably improves the prediction results (Table~\ref{table:combination}) across all chains. 
			For example when using supervised 
			learning in the case of Starbucks an NDCG@$10$ score of $0.77$ is reached an improvement in  
			the performance of Transition Quality, our best single feature prediction, 
			by $10\%$. In the Dunkin' Donuts and McDonald's cases the best NDCG@$10$ in 
			supervised learning reaches $0.83$ and $0.84$ in comparison to the best single feature performance that achieves 
			$0.81$ and $0.79$, respectively. 		
			The improvement in performance when supervised learning is applied is more clearly indicated when 
			taking into account the Accuracy@X\% measurement.
			The top predicted geographic location for Starbucks, as shown in Figure~\ref{fig:combination}, 
			is positioned with an accuracy of $67\%$ and $76\%$ amongst the Top-$10$\% and the Top-$15$\% in the list $\overline{R}$ of the most popular places. 
			When considering the Top-$20$\% most popular places, accuracy goes over the mark of $80$\% across all considered chains, reaching 
			a maximum of $93\%$ in the case of Dunkin' Donuts. 
			These results significantly outperform the best individual features 
			predictions which are $58\%$ for Starbucks, $75\%$ and $81\%$ for McDonald's and Dunkin' Donuts respectively. 
		
			Overall, the combination of geographic and mobility features in a supervised learning algorithm 
			yields better results than the single feature predictions. Moreover, when considering mobility features in 
			addition to the geographic features we have observed large improvements, highlighting that information about 
			the check-in movements of Foursquare users adds value to the prediction. 
			Across different prediction techniques, we have observed that the optimal performance is achieved for the predictions based 
			on the Support Vector Regression algorithm.

\section{Implications}
\label{sec:discussion}
The analysis and evaluation presented in the previous sections highlight how the new generation of location-based services 
can play a significant role in the commercial evolution of the web. 
The rich spatio-temporal datasets sourced from services 
like Foursquare can provide access to new layers of data bringing us one step
towards the integration of local businesses in the economic model of social networking platforms. 
As we have seen in our experiments, knowledge on the semantics of geo-tagged venues can provide effective data representations to model the commercial value of urban areas allowing us to measure the influence of competitive retail facilities nearby or the presence of types of venues that can generate a large flow of customers for a target business. 

The fact that different features may prove more or less effective across chains signifies that the problem of optimal retail store placement is not trivial. 
Different types of businesses can demonstrate significant variations in the ways they become attractors to their respective customer crowds. As empirically
studied in Section~\ref{sec:analysis} the spatial properties and the patterns of customer movement can be chain specific, in spite of the fact that common patterns of attraction are also observed. As we have shown, supervised learning classifiers can exploit a number of data mining features and seamlessly alleviate the biases due to heterogeneities in the way retail facilities attract users.
Moreover, the combination of geographic and mobility features in the majority of cases has yielded superior performance suggesting that the dynamics of human movement 
matter in understanding the retail quality of an area. 

Besides, is especially clear in the case of Starbucks, noise can impede the prediction task. One factor that has contributed to this is the fact that very proximate 
retail facilities of the same chain will have by definition very similar feature values, although, at the same time, their popularity may vary significantly. Other factors such as local architecture
and planning idiosyncrasies of an area can effect the flow of customers towards a place. Especially, in extremely dense cities, like New York, two venues can have similar latitude and longitude coordinates
but one of them may be placed at the corner of a high street and the other at the top of a skyscraper nearby. Along these lines, we are planning to extend our work to include more cities and chains. 
Informal experiments we have conducted suggest the existence of large heterogeneities in the spatial properties of different cities, but also a strong biases in the amount of user check-ins 
from area to area. Therefore, the development of techniques that deal with these issues is a challenging task with potentially high value for location-based services and urban data mining in general.\newline \newline

\section{Related work}
\label{sec:related}
The retail store placement problem has, in recent decades, attracted researchers
from a broad spectrum of disciplines. 
Land economy community research has concentrated on 
spatial interaction models, which are based on the assumptions that the intensity of interaction 
between two locations decreases with their distance and that the usability
of a location increases with the intensity of use and the proximity 
of complementary arranged locations~\cite{athiyaman2010,berman2002,kubis2007}.  
It was shown, however, that the applicability of these models is limited to agglomerations, such as big 
shopping centres, and their predictive accuracy decreases when smaller, specialized stores are considered.  
With respect to previous work in the general area, in this paper we examined how the problem 
can be framed in terms information signals mined from location-based services. As we have seen, the richness
of information provided in these services could help us to study the retail quality of an area in a fine grained manner: various types of geographic, semantic and mobility information not only can complement traditional
techniques, but also form the basis for a new generation of business analytics driven by services such
as Foursquare.  

The present paper has been largely inspired by Jensen's seminal work~\cite{jensen2006,jensen2009} on the
identification of the appropriate geographic positioning of retail stores. Jensen's approach
uses a spatial network based formulation of the problem, where nodes are $55$ different
types of retail stores and weighted signed links are defined to model the attraction and repulsion 
of entities in the network. Subsequently, a retail quality index is devised which 
has been used to empirically assess the positioning of stores in the city of Lyon, France. 
Another approach based on the analysis of the spatial distribution of commercial activities was 
proposed by Porta et al. in~\cite{porta2011,porta2009}. The authors investigate the relationship between street centrality and retail
store density in the cities of Bologna and Barcelona respectively, verifying how the former acquires a significant role 
in the formation of urban structure and land usage. We extend the results of these works by adding to the analysis 
features mined from the human mobility traces and effectively show that the combination of the geographic and mobility features 
provides better insights on the quality of an area as a potential spot to open a new retail facility. To our knowledge
this is also one of the first efforts to tackle the problem exploiting machine learning algorithms. These techniques may prove
crucial in doing similar analyses in the future, as they have been devised to operate in settings where large amounts of dynamic data are
available.

Finally, from a data mining perspective, we could classify our work in the area of \textit{urban mining}, which studies the extraction of knowledge  from spatial or geographic datasets and aims to improve services and intelligence in the city. In~\cite{Quercia_2010}, Quercia et al. mine cellular data of user movements in the city of Boston to recommend social events. In~\cite{Lathia:2011} the authors analyze the movement of passengers of the London metro and provide insights into the financial spending of transport users, and in~\cite{Lathia:2012} the relationship between social deprivation on citizen mobility is investigated. In~\cite{yuan2012discovering} the authors attempt to infer the functions of different 
regions in the city by analyzing spatial distribution of commercial activities and human mobility traces in the city of Beijing. 
The present paper is well aligned with this stream of work and extends the applicability of the urban mobility mining methods to the field of retail analytics.

\section{Conclusion}
\label{sec:conclusions}
In this paper we tackled the problem of 
optimal retail store placement in the context of location-based 
social networks. We collected  human 
mobility data from the leading location-based service, 
Foursquare, and analyzed it to understand how the popularity 
of three retail store chains in New York is shaped,
in terms of number of check-ins.

We developed and evaluated a diverse set of data mining features, 
modeling spatial and semantic information about places and patterns of user movements 
in the surrounding area. 
We evaluated each feature separately and found that, 
among those exploiting place semantics, the presence of user attractors (i.e., train station or airport) 
as well as retail stores of the same type to the target chain (i.e. coffee shop or restaurant) encoding 
the local commercial competition of an area, 
are the strongest indicators of popularity. However, additional improvement
in the prediction performance may be achieved by assessing potential flows of users that 
a place may attract from other venues nearby or far away locations. 
We further combined different features in a set of 
supervised learning algorithms and showed that the
popularity of places can be better explained by the fusion of geographic and 
mobility features. 
We plan to extend our work to the comparison of the impact 
of features in different cities and in new types of places and retail chains.
\newline

\section{Acknowledgements}
We acknowledge the support of EPSRC through grants MOLTEN (EP/I017321/1)
and GALE (EP/K019392).

\bibliographystyle{plain}
{\small

}

\end{document}